# Connectome brain fingerprinting: terminology, measures, and target properties

Matteo Fraschini, Matteo Demuru, Daniele Marinazzo, and Luca Didaci

*Abstract*—Distinguishing one person from another—what biometricians call recognition —is extremely relevant for different aspects of life. Traditional biometric modalities (fingerprint, face, iris, voice) rely on unique, stable features that reliably differentiate individuals. Recently, the term "fingerprinting" has gained popularity in neuroscience, with a growing number of studies adopting the term to describe various brain-based metrics derived from different techniques. However, we think there is a mismatch between its widely accepted meaning in the biometric community and some brain-based metrics. Many of these measures do not satisfy the strict definition of a biometric fingerprint— that is, a stable trait that uniquely identifies an individual. In this study we discuss some issues that may generate confusion in this context and suggest how to treat the question in the future. In particular, we review how "fingerprint" is currently used in the neuroscience literature, highlight mismatches with the biometric-community definition, and offer clear guidelines for distinguishing genuine biometric fingerprints from exploratory similarity metrics. By clarifying terminology and criteria, we aim to align practices and facilitate communication across fields.

*Index Terms*—Brain fingerprinting, identification, verification.

## I. Introduction

Finding what makes humans "unique" or at least distinguishable from others is extremely relevant for different aspects of life. Recently, the need to develop systems for acquiring and storing personal characteristics and performing automatic recognition has emerged [1], [2], [3], [4].

Matteo Fraschini and Luca Didaci are with the Department of Electrical and Electronic Engineering, University of Cagliari (e-mail: matteo.fraschini@unica.it; didaci@unica.it). Matteo Demuru is with Liceo Scientifico e Linguistico Statale "G. Marconi", 07100 Sassari, Italia (e-mail: matteo.demuru@scuola.istruzione.it). Daniele Marinazzo is with Department of Data Analysis, Ghent University, 1 Henri Dunantlaan, B-9000 Ghent, Belgium (e-mail: daniele.marinazzo@ugent.be).

To date, there are a myriad of different applications that make use of such an approach in daily life. In this context, the term "biometrics" has been used to include all the methods aiming at automated recognition of human individuals, such as dactyloscopic fingerprint, face, iris, and voice. The dactyloscopic fingerprint, produced by the pattern of ridges and valleys on the surface of a fingertip, has been used in forensic applications for over 100 years and represents the most widely used method for personal recognition.

In 2015, a paper titled "Functional connectome fingerprinting: identifying individuals using patterns of brain connectivity" [5] was published in Nature Neuroscience. Since then, the word fingerprinting has gained popularity in neuroscience, with an increasing number of studies investigating it. Before discussing how these studies contributed to the field, it is essential to take a step back in the past and understand the origin of the terminology and its meaning. According to the Oxford English Dictionary, the term fingerprinting refers to "the action or practice of taking a person's fingerprints, […]" where the fingerprint is "an impression or mark made on a surface by the tip of a person's finger […]", or "something that identifies" according to Merriam-Webster. Back to the paper by Finn et al, in this case, the attempt to use the functional connectome (the complete set of statistical dependencies between recordings in different brain regions) derived from functional magnetic resonance imaging (fMRI) as an impression or mark to identify individuals emerges. The authors argue that "an individual's functional brain connectivity profile is both unique and reliable, similarly to a fingerprint". This is not surprising since, historically, (dactyloscopic) fingerprints represent the most common biometric characteristic to perform recognition (i.e., establish a person's identity).

The focus of several applications in neurosciences has then moved to the question "how different is a subject from the other subjects?", while keeping the "fingerprinting" name. This is a different question, yet the limits have been kept blurry, introducing some confusion due to the mismatch with what is already known in the biometrics community about the methods used to approach recognition. Here, we discuss some issues that may generate confusion in this context and suggest how to treat the question in the future.

## II. Biometric functionalities and errors

A biometric system is a pattern recognition system that acquires data from an individual, extracts a set of features, and compares these features with a template in the database. It is relevant to highlight two main functionalities when dealing with recognition: verification and identification. This distinction is not irrelevant since the two functionalities may need different methods to measure the performance. By contrast, the word recognition is usually used as a more general term without a specific reference to one of the two functionalities.

In verification mode, the system confirms or denies a claimed identity. The goal is to verify a claimed identity, so the comparison is made only against the templates corresponding to the claimed identity (one-to-one matching). The system must answer the question: ''Does this biometric data belong to Bob?''.

In identification mode, the system determines an individual's identity among a set of known individuals. The comparison is made against templates of all enrolled users (one-to-many matching). The system must answer the question: ''Whose biometric data is this?''.

It is important to note that the terms recognition, identification, verification, and authentication have sometimes been used incorrectly and ambiguously in the literature, see [4], [6] for a brief description. The main module of a biometric system is the "matcher module", where the features from the subject involved in the recognition are compared with the features collected in the dataset, and the so-called matching scores are obtained. The distribution of matching scores obtained by comparing samples from the same person is called the genuine distribution. In contrast, the distribution of matching scores obtained by comparing samples from different persons is called the impostor distribution. The basic accuracy measures for biometric verification are the False Accept Rate (FAR) and the False Reject Rate (FRR), which measure the proportion of successful or failed matching attempts using genuine and impostor samples. Given a set of genuine and impostor scores, FRR and FAR may assume different values depending on the defined threshold; if the threshold increases, the FAR decreases while the FRR increases, and vice versa. The list of FAR and FRR values, as the threshold varies, fully describes the performance of the biometric verification system, since these metrics estimate the probability that, for a predefined threshold, an impostor will be accepted as genuine (FAR) or, conversely, that a genuine individual will not be recognized (FRR). This information can be represented graphically via the Receiver Operating Characteristic (ROC) curve, which plots the FRR against the FAR. Many ways are possible to obtain a single summary performance value, for example the area under the ROC curve (AUC) or the value of the system error in a specific point of the ROC curve. Typical examples of those indexes are the Equal Error Rate (EER), which denotes the system error when FRR = FAR, and the ZeroFAR, which denotes the FRR values when FAR = 0. For an identification system, the most commonly used metric is the rank-k identification rate (Rk), which is the number of times the correct identity appears within the top k most likely candidates. The Cumulative Match Characteristic (CMC) curve plots Rank-performance against k and provides a visual representation of the system's performance across different ranks. Therefore, if a user's correct identity corresponds to the highest score among all the match scores, the user is identified as rank one. Other standard measures are the True Positive Identification Rate (TPIR), which represents the proportion of correctly identified individuals out of the total number of individuals who should be correctly identified, and the Correct Recognition Rate (CRR), which represents the proportion of correctly recognized individuals out of the total number of individuals presented in the system.

Any performance measure for a biometric system must be directly tied to its intended purpose—verification or identification—and thus reflect the corresponding success and failure rates. Those rates depend on the shape and overlap of the genuine and impostor score distributions, as well as on the threshold(s) that the system adopts. FAR, FRR, and rank-k each directly capture how often the system succeeds or fails in its specific task. Any new proposal for measuring "distance between individuals" must preserve an explicit, operational link to these success/failure rates. Without that link—i.e., without reference to how genuine and impostor scores separate under the chosen threshold(s)—a proposed metric cannot meaningfully characterize biometric performance.

## III. Brain Fingerprinting

In the last few years, "brain fingerprinting" has become a relevant topic in neuroscience. The fingerprint is often assumed to be the "functional connectivity" (FC), a term indicating both the measure (statistical dependency, often correlation) and its target property (the correlated neural activity). Nevertheless, other fingerprints have been proposed in the literature [7], [8], [9], [10]. Despite the well-known metrics described in section 2, other measures have been proposed to evaluate identification performance, this time in terms of differentiability.

After the seminal study on FC fingerprinting [5], which applied a standard measure (rank-k identification rate with k = 1), in 2018, Amico et al. [11] proposed a framework to assess individual fingerprint and to maximize its identifiability of functional connectomes where differential identifiability ($I_{diff}$) is introduced. The choice of this measure was motivated as the quest for "a more continuous score on the level of individual fingerprinting present on a set of test-retest functional connectomes". Later, da Silva Castanheira et al. [12] extended the $I_{diff}$ and proposed a related differentiability measure, $D_{self}$, computed at the subject level, with the purpose of assessing how easy or difficult it is to differentiate one participant from the rest of the group. In more details, differential identifiability ($I_{diff}$) quantifies the difference between the average within-subject similarity and the average between-subjects similarity from an identifiability matrix containing the correlations between the subjects' functional connectivity at the test and retest. $I_{diff}$ is computed as follows:

$$I_{diff} = (I_{self} - I_{others}) * 100 \qquad (1)$$

where $I_{self}$ is the average of the main diagonal elements of the identifiability matrix and $I_{others}$ is the average of the off-diagonal elements.

The authors claim that "The higher the value of $I_{diff}$, the higher the individual fingerprint overall along the population". $D_{self}$ was proposed as an extension of $I_{diff}$, and it was defined as the z-score of participant Pi's correlation to themselves between two different time points (or datasets), with respect to Pi's correlation to all other individuals in the cohort. $D_{self}$ is computed as follows:

$$D_{self} = (Corr_{ii} - \mu_{ij})/\sigma_{ij} \quad (2)$$

where $Corr_{ii}$ represents the correlation between the measures of subject Pi's at the two time points, $\mu_{ij}$ the mean correlation between the measure of participant Pi at first time point and those of all other individuals at the second time point, and $\sigma_{ij}$ the standard deviation. The authors claim that if a participant is easily differentiable, its related $D_{self}$ increases.

## IV. EXPERIMENTAL RESULTS

Here, we show the relationship between these measures and the metrics originally introduced to estimate the performance of a biometric system. To test this relationship, we simulated identifiability matrices using the Beta distribution (defined on the interval [0, 1] and is parameterized by two positive parameters, alpha and beta), which allows controlling the generated distributions' shape. Alpha and beta parameters were derived from different mean and standard deviation values to define a total of 256 different scenarios. The only constraint was forcing the mean of the diagonal elements (genuine scores) to be higher than that of the out-diagonal elements (impostor scores). An identifiability matrix was obtained for each of the 256 scenarios, and EER, $I_{diff}$ and $D_{self}$ were computed.

All the code to perform the simulations is available at the following link: https://github.com/matteogithub/Idiff_ERR.

Figure 1 shows the corresponding scatterplot between EER and $I_{diff}$, where Pearson's correlation coefficient is -.707.

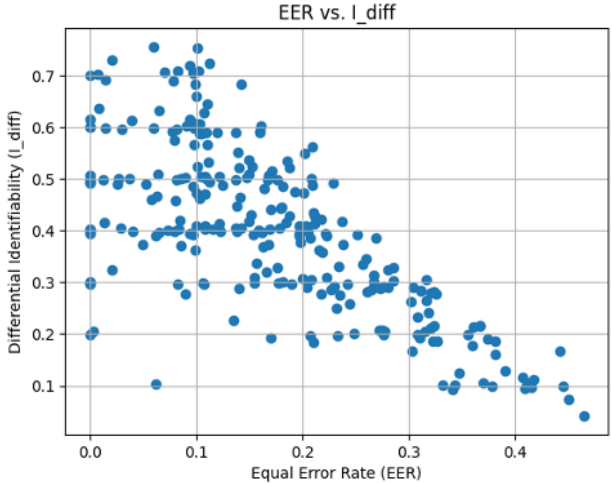

Fig. 1. Scatterplot showing the relationship between EER and $I_{diff}$ for different simulated identifiability matrices.

Despite the high correlation, it is evident that for values of EER

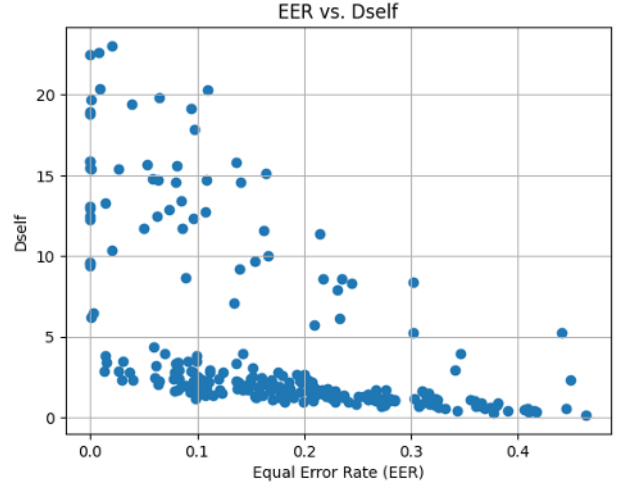

Fig. 2. Scatterplot showing the relationship between EER and $D_{self}$.

equal to 0, $I_{diff}$ varies over a large range of values.

The same approach was then used to explore the association between EER and $D_{self}$ (averaged over all the subjects) for different beta distributions. In this case, the Pearson's correlation coefficient was -.536, and Figure 2 shows the corresponding scatterplot. Again, despite the reported good correlation, it is still evident that for values of EER equal to 0, $D_{self}$ varies over a large range of values. The distribution of $I_{diff}$ vs EER seems heteroscedastic, with the former varying over a large range for values of EER approaching zero.

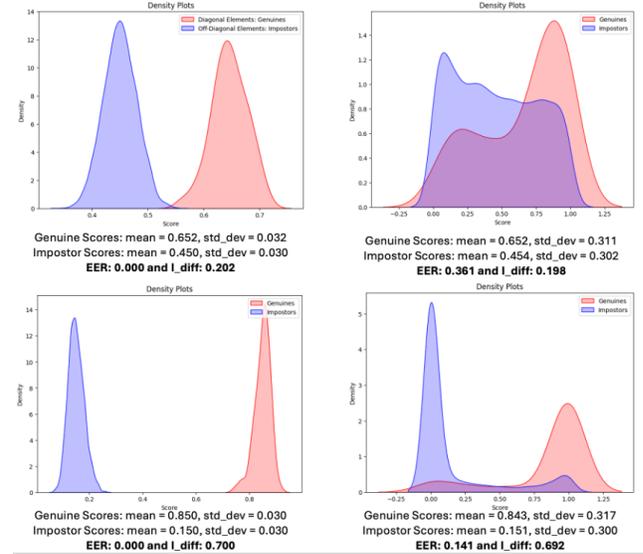

Fig. 3. Density plots showing the genuine and impostor scores distributions for four different scenarios.

In Figure 3, we show in more details the relationship between EER and $I_{diff}$, reporting the underlying distributions where this discrepancy emerged. In particular, the two upper panels of Figure 3 show that for two very different identification scenarios, an almost optimal scenario (EER approaches 0) as depicted in the left panel, or a very poor scenario (EER is much higher) where the distributions have a significant overlap, $I_{diff}$ wrongly provides a low, yet similar value. Moreover, the two lower panels of Figure 3 show that, again, for two very different identification scenarios, an almost optimal scenario (EER approaches 0) as depicted in the left panel, or a poor scenario (EER is much higher) where the distributions have a significant overlap, $I_{diff}$ assumes a high, yet similar value.

## V. Conclusion

In this work, we summarized the definition of fingerprinting, remarked the biometrics functionalities and their evaluation metrics, and showed the relationship between some measures ($I_{diff}$ and $D_{self}$) [11], [12] recently proposed to quantify the "identifiability capability" and the EER (an established and common metric used to summarize the performance of a biometric system in a verification problem). This relationship was explored using simulated identifiability matrices from Beta distributions parameterized by varying the alpha and beta parameters, which allow to control the shape of the distributions. Despite the high correlations observed, the results show that both $I_{diff}$ and $D_{self}$ may provide a wide range of values, from very high to very low, for optimal scenarios where correctly EER approaches 0. Moreover, the same issue emerged for poor outcome scenarios where EER tends to be higher, and both $I_{diff}$ and $D_{self}$ values span over a large range.

$I_{diff}$, by construction, provides an estimate of the distance between the average of the main diagonal elements ($I_{self}$ or genuine scores) of the identifiability matrix and the average of the off-diagonal elements ($I_{others}$ or impostor scores) but does not consider the overlap between the two distributions. This problem could lead to rating a system as very good (high identifiability) even when, in fact, it is not, and on the other hand, provide a low score value (low identifiability) in a scenario where recognition is potentially very high.

The EER in verification represents a robust way to describe the behavior of a biometric system, which essentially depends on the shape of the score distributions of Genuine and Impostors and the threshold adopted. It is worth noting that an EER of 0 indicates that both the False Accept Rate (FAR) and False Reject Rate (FRR) are 0, meaning the impostor and genuine distributions are perfectly separated. In this case, a threshold can be found that allows for error-free identification within the biometric system. However, the separation between these distributions depends not only on the distance between their mean values but also on factors such as their dispersion, skewness, and overall shape. This explains why, even with an EER of 0, $I_{diff}$ and $D_{self}$ can vary widely. The distance between the means of these two distributions (as computed by $I_{diff}$) does not characterize the performance of a biometric system.

Including an analysis based on real data seems superfluous in this context. $I_{diff}$, differently from EER, does not have a reference value for considering identification high or low, so any result deriving from a real dataset would not be easily interpretable. It is evident from the reported simulation that it is not possible to exclude cases where $I_{diff}$ and EER point in the same direction, however, there are several other cases where this does not happen. Based on the distributions of the real values, we will find ourselves in one of the possible cases outlined by the different distributions reported by the simulation, while our point here is general.

Finally, $I_{diff}$ is derived from the matrix containing the correlation values, so any dependence from a specific acquisition system (i.e., EEG, MEG or fMRI) used to measure brain activity looks irrelevant in this context. We thus suggest using consolidated measures, such as rank-k identification rate or EER, to report the performance of identification/verification when the goal is indeed fingerprinting or identifiability, and other measures when the question is a different one.

Recently, several studies [13], [14], [15], [16] have adopted $I_{diff}$ or $D_{self}$ in contexts that deviate significantly from their original intended purpose. Here, the concept of "fingerprinting" has been applied broadly to scenarios that merely involve comparisons between groups, without clear evidence of identification capabilities. Such an approach misrepresents the foundational idea of fingerprinting, which explicitly implies the ability to uniquely and accurately distinguish one entity from another. This has been recently discussed and recognized in a different domain [17]. However, even in contexts oriented towards individual identification, the use of $I_{diff}$ and $D_{self}$ alone is problematic. These metrics, derived from mean differences in within-subject and between-subject correlation values, do not adequately capture the crucial aspects of distribution overlap and score variability. Consequently, they poorly reflect true recognition performance, as clearly demonstrated by their inconsistent relationship with robust identification metrics such as the Equal Error Rate (EER).

In conclusion, we should avoid using the term 'fingerprinting' when the investigation does not relate to the identification of a unique and distinguishable feature that can differentiate one entity from another. Specifically, calculating the difference between the distributions of correlation values within a single subject and between different subjects (without considering recognition errors) is not related to 'fingerprinting.' This is especially true when these differences are derived simply by subtracting the mean values of the two distributions. This is not a matter of trivial terminological precision but a fundamentally important issue to align with the meaning consistently used across other scientific fields, where 'fingerprinting' is closely associated with the ability to uniquely and accurately identify entities (i.e., device fingerprint, genetic fingerprinting, chemical fingerprint).

# REFERENCES

[1] A. K. Jain, P. Flynn, e A. A. Ross, *Handbook of Biometrics*. Springer Science & Business Media, 2007.

[2] Q. Gui, M. V. Ruiz-Blondet, S. Laszlo, e Z. Jin, «A Survey on Brain Biometrics», *ACM Comput. Surv.*, vol. 51, fasc. 6, p. 112:1-112:38, feb. 2019, doi: 10.1145/3230632.

[3] A. A. Ross, K. Nandakumar, e A. K. Jain, *Handbook of Multibiometrics*. in International Series on Biometrics. Springer US, 2006. doi: 10.1007/0-387-33123-9.

[4] J. L. Wayman, «Biometric Verification/Identification/Authentication/Recognition: The Terminology», in *Encyclopedia of Biometrics*, S. Z. Li e A. Jain, A c. di, Boston, MA: Springer US, 2009, pp. 153–157. doi: 10.1007/978-0-387-73003-5_206.

[5] E. S. Finn *et al.*, «Functional connectome fingerprinting: identifying individuals using patterns of brain connectivity», *Nature Neuroscience*, vol. 18, fasc. 11, pp. 1664–1671, nov. 2015, doi: 10.1038/nn.4135.

[6] A. K. Jain, A. Ross, e S. Prabhakar, «An introduction to biometric recognition», *IEEE Transactions on Circuits and Systems for Video Technology*, vol. 14, fasc. 1, pp. 4–20, gen. 2004, doi: 10.1109/TCSVT.2003.818349.

[7] M. Demuru e M. Fraschini, «EEG fingerprinting: subject-specific signature based on the aperiodic component of power spectrum», *Computers in Biology and Medicine*, vol. 120, pp. 1–5, 2020, doi: 10.1016/j.compbiomed.2020.103748.

[8] D. La Rocca *et al.*, «Human Brain Distinctiveness Based on EEG Spectral Coherence Connectivity», *IEEE Transactions on Biomedical Engineering*, vol. 61, fasc. 9, pp. 2406–2412, set. 2014, doi: 10.1109/TBME.2014.2317881.

[9] M. DelPozo-Banos, C. M. Travieso, C. T. Weidemann, e J. B. Alonso, «EEG biometric identification: a thorough exploration of the time-frequency domain», *J Neural Eng*, vol. 12, fasc. 5, p. 056019, ott. 2015, doi: 10.1088/1741-2560/12/5/056019.

[10] H.-L. Chan, P.-C. Kuo, C.-Y. Cheng, e Y.-S. Chen, «Challenges and Future Perspectives on Electroencephalogram-Based Biometrics in Person Recognition», *Front Neuroinform*, vol. 12, ott. 2018, doi: 10.3389/fninf.2018.00066.

[11] E. Amico e J. Goñi, «The quest for identifiability in human functional connectomes», *Sci Rep*, vol. 8, fasc. 1, Art. fasc. 1, mag. 2018, doi: 10.1038/s41598-018-25089-1.

[12] J. da Silva Castanheira, H. D. Orozco Perez, B. Misic, e S. Baillet, «Brief segments of neurophysiological activity enable individual differentiation», *Nat Commun*, vol. 12, fasc. 1, p. 5713, set. 2021, doi: 10.1038/s41467-021-25895-8.

[13] J. da S. Castanheira *et al.*, «The neurophysiological brain-fingerprint of Parkinson's disease», *eBioMedicine*, vol. 105, lug. 2024, doi: 10.1016/j.ebiom.2024.105201.

[14] S. Stampacchia *et al.*, «Fingerprints of brain disease: connectome identifiability in Alzheimer's disease», *Commun Biol*, vol. 7, fasc. 1, pp. 1–16, set. 2024, doi: 10.1038/s42003-024-06829-8.

[15] M. Ambrosanio *et al.*, «The Effect of Sleep Deprivation on Brain Fingerprint Stability: A Magnetoencephalography Validation Study», *Sensors*, vol. 24, fasc. 7, Art. fasc. 7, gen. 2024, doi: 10.3390/s24072301.

[16] L. Cipriano *et al.*, «Brain fingerprint and subjective mood state across the menstrual cycle», *Front. Neurosci.*, vol. 18, dic. 2024, doi: 10.3389/fnins.2024.1432218.

[17] J. W. Lockhart, A. Fuentes, G. Rippon, e L. Eliot, «Not so binary or generalizable: Brain sex differences with artificial neural networks», *Proceedings of the National Academy of Sciences*, vol. 122, fasc. 2, p. e2411917121, gen. 2025, doi: 10.1073/pnas.2411917121.